\documentclass[aps,prb,preprint,amsmath,amssymb,superscriptaddress,longbibliography,floatfix,12pt]{revtex4-1}

\usepackage{graphicx, multirow, xcolor, bm, ulem}
\usepackage{amsfonts,amssymb,amsmath}
\usepackage{graphicx,dcolumn,bm,color,braket,slashed}
\usepackage{times, comment, mathtools, extarrows,textcomp,ulem}

\def\ket#1{|#1\rangle }
\def\bra#1{\langle #1 |}

\begin{document}
\title{Surface $p$-wave Superconductivity and Higher-Order Topology in MoTe$_2$}

\author{Sangyun Lee}
\thanks{These authors contributed equally: Sangyun Lee and Myungjun Kang}
\affiliation{Center for Quantum Materials and Superconductivity (CQMS) and Department of Physics, Sungkyunkwan University, Suwon 16419, South Korea}
\affiliation{Los Alamos National Laboratory, Los Alamos, NM 87545, USA}
\affiliation{Department of Physics and National High Magnetic Field Laboratory, University of Florida, Florida 32611, USA}

\author{Myungjun Kang}
\thanks{These authors contributed equally: Sangyun Lee and Myungjun Kang}
\affiliation{Department of Physics, Hanyang University, Seoul 04763, South Korea}

\author{Duk Y. Kim}
\affiliation{Agency for Defense Development, Daejeon 34186, South Korea}

\author{Jihyun Kim}
\affiliation{Center for Quantum Materials and Superconductivity (CQMS) and Department of Physics, Sungkyunkwan University, Suwon 16419, South Korea}

\author{Suyeon Cho}
\affiliation{Division of Chemical Engineering and Materials Science, Ewha Womans University, Seoul, 03760, South Korea}

\author{Sangmo Cheon}
\email[Corresponding authors:]{sangmocheon@hanyang.ac.kr}
\affiliation{Department of Physics, Hanyang University, Seoul 04763, South Korea}
\affiliation{Research Institute for Natural Science and High Pressure, Hanyang University, Seoul 04763, South Korea}

\author{Tuson Park}
\email[Corresponding authors:]{tp8701@skku.edu}
\affiliation{Center for Quantum Materials and Superconductivity (CQMS) and Department of Physics, Sungkyunkwan University, Suwon 16419, South Korea}

\begin{abstract}
Exploration of nontrivial superconductivity and electronic band topology is at the core of condensed matter physics and applications to quantum information.
The transition-metal dichalcogenide (TMDC) MoTe$_2$ has been proposed as an ideal candidate to explore the interplay between topology and superconductivity, but their studies remain limited regarding the required high-pressure environments.
Here, we observe proximity-induced surface $p$-wave superconductivity, and investigate the higher-order topological nature of MoTe$_2$ in its 1T$'$ phase, which emerges from the T$_d$ phase through a high-pressure-induced topological phase transition.
Using surface-sensitive soft-point-contact Andreev reflection spectroscopy, we confirm the emergence of surface $s+p$-wave superconductivity via the BTK model as well as a zero-bias conductance peak.
Such surface $p$-wave superconductivity emerges via the proximity effect between an $s$-wave superconducting band and a second-order topological band, which is protected by the time-reversal and inversion symmetries.
The temperature dependence of the surface $p$-wave superconducting gap shows a correlation with that of the bulk $s$-wave gap, as well as its suppression by an external magnetic field or a reduction in pressure, implying its proximity-induced origin.
Moreover, we suggest that the topological hinge states, derived from second-order topological bands, evolve into zero-energy Majorana corner states in this proximity-effect-induced third-order topological superconducting phase.
These results demonstrate the potential realization of topological superconductivity in MoTe$_2$, thus opening a pathway for studying various topological natures of TMDC materials.
\end{abstract}

\maketitle

\begin{figure}[t]
\includegraphics[width=0.8\textwidth]{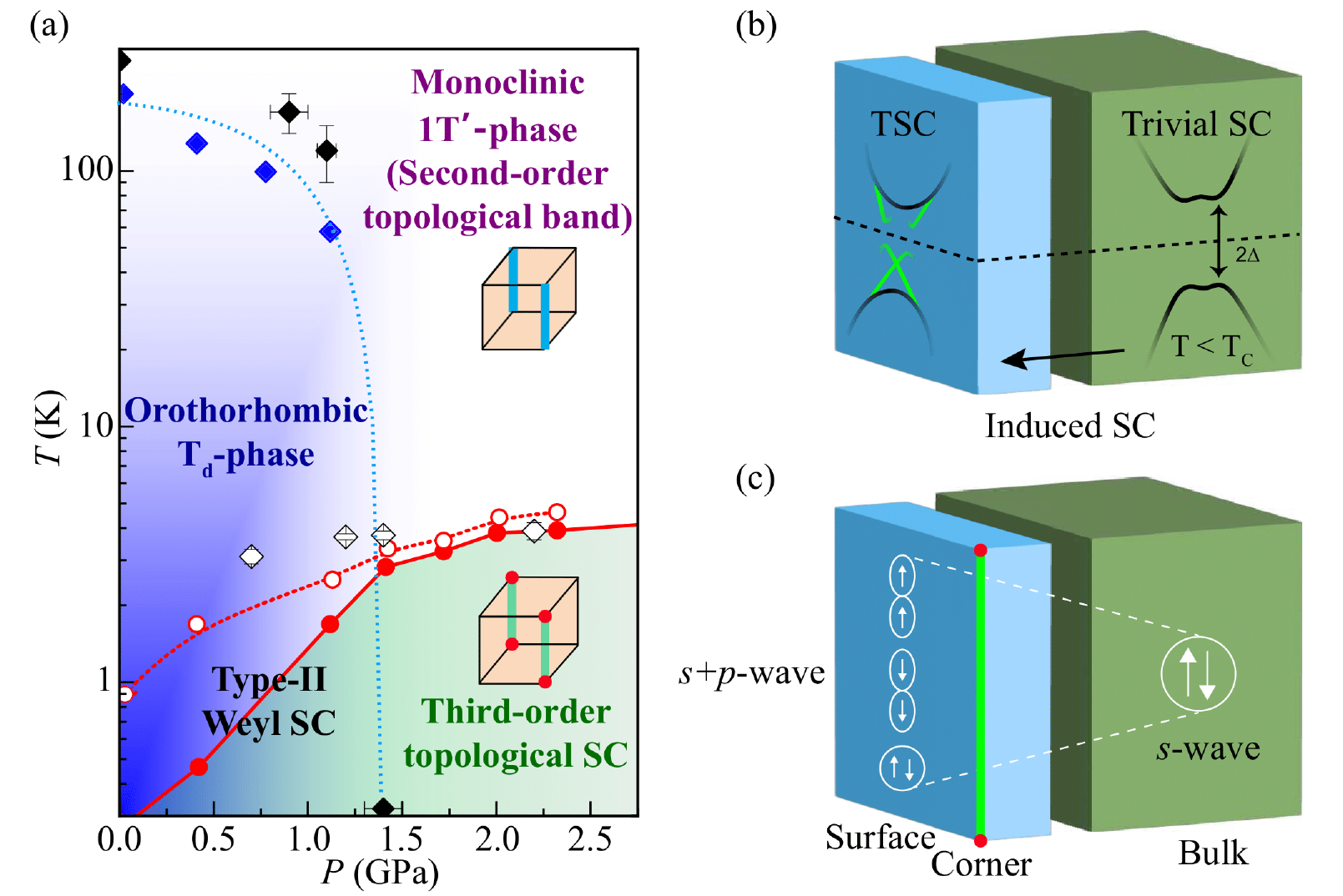}
\caption{\label{fig1:crys_struct} 
\textbf{(a)} Temperature–pressure phase diagram of MoTe$_2$.
The normal 1T$'$ phase hosts second-order topological insulating bands, and its SC phase is a third-order topological superconductor.
Solid blue and black diamonds indicate the structural phase transition temperature ($T^*$) from 1T$'$ to T$_d$ phases determined from the electrical resistivity and X-ray scattering studies, respectively~\cite{lee2018origin,guguchia2020structure}.
Open and solid red circles indicate the onset temperature of the SC phase transition ($T_{\text{c, onset}}$), and the zero-resistance temperature ($T_{\text{zero}}$) observed in the transport experiment, respectively~\cite{lee2018origin}. 
Open black diamonds indicate the SC transition temperature ($T_{c}$) obtained from our soft-point contact Andreev reflection spectroscopy, which comparably follows $T_{\text{c, onset}}$ measured in the transport measurement.
The suppression of $T^*$ coincides with a sharp increase of $T_c$ in the T$_d$ phase below the quantum critical point ($P_c \approx 1.4$~GPa).
\textbf{(b)} 
Schematics for the bulk-to-surface proximity effect between the $s$-wave bulk superconductivity and helical Dirac-type surface states generated from the SOTI bands. 
Below $T_c$, the $s$-wave superconducting state in the bulk is gapped.
Simultaneously, the bulk-to-surface proximity effect induces gapped hinge states, the dispersions of which are given in green.
%
\textbf{(c)} 
Schematics for the emergence of $s+p$-wave pairing on the surface via bulk-to-surface proximity effect between the $s$-wave bulk superconductivity and helical Dirac-type surface states.
In (a,c), for the third-order topological superconducting phases, zero-energy Majorana corner states (red) are located at the end of the hinges (green). 
}
\end{figure}

\clearpage
\section{Introduction} \label{sec:intro}
Nontrivial electronic band topology and superconductivity are at the forefront of condensed matter physics, with significant implications for quantum information technologies.
Notable examples include topological insulators, crystalline insulators, Dirac/Weyl semimetals, and, in particular, topological superconductors, which are predicted to host Majorana states obeying non-Abelian statistics~\cite{Hasan2010rev,Fu2011tci,Sato2017tsc}.
Recently, the concept of topology has been extended to materials with higher-order topology, where a $d$-dimensional $n$th-order topological material has topologically protected surface states at its $(d-n)$-dimensional surface~\cite{schindler2018higher}.
Although such higher-order systems have been reported in photonic systems~\cite{Ahmed2023photo,Zhu2019poto}, their condensed matter counterpart is still elusive~\cite{choi2020wte2,Sato2017tsc,kallin2016chiral,Fu2010bise,sato2013inte}.

Transition metal dichalcogenides (TMDCs) with van der Waals bonding have drawn considerable attention because they can host nontrivial topology and superconductivity~\cite{wang2019higher,lane2022identifying,schrade2024nematic,xu2014topological,bahramy2018ubiquitous}. 
Among these, MoTe$_2$ stands out as a promising candidate to study the interplay between topology and superconductivity because of its anticipated novel topological states~\cite{qian2014quantum,keum2015bandgap,
deng2016experimental, zhang2017experimental, ma2019experimental,  Hsu2020wtesc, Jahin2023wteto, Zeng2023motechern, liu2020quantum, Heikes2018struture}, as illustrated in Fig.~\ref{fig1:crys_struct}.
At ambient pressure, the T$_d$ phase exhibits Fermi arcs arising from the Weyl fermion~\cite{tamai2016fermiarc,laia2018stm} and it is suggested that below the superconducting critical temperature ($T_c$) topological superconductivity with $s^\pm$-wave pairing emerges~\cite{guguchia2017signatures, naidyuk2018surface}.
%
At higher pressure, MoTe$_2$ in the 1T$'$ phase is suggested to host higher-order topological bands, with time-reversal and inversion symmetries protecting the topological hinge states~\cite{wang2019higher,ezawa2019second,su2022persistence,pan2023altshuler,chen2024asymmetric,huang2024hybrid}.

From the viewpoint of the implementation of topological superconductivity, this system can be advantageous for the following reasons.
One prototypical method to realize topological superconductivity is via $p$-wave superconductivity, which has two plausible routes.
The first is for the $p$-wave superconductivity to be intrinsic, prominent candidates for this method are Sr$_2$RuO$_4$ and Cu$_x$Bi$_2$Se$_3$~\cite{sasaki2011topological,kallin2012chiral}.
However, the superconductivity is weak to disorder and there are only a few potential materials. 
The other method is external, i.e. the proximity-induced $p$-wave superconductivity, which is achieved by a hetero-structure of a topological insulator and an $s$-wave superconductor, resulting in a topological superconductor at the interface~\cite{fu2008superconducting,lutchyn2010majorana}.
However, a long superconducting (SC) coherence length is required as well as a complicated hetero-structure to realize $p$-wave superconductivity.
On the other hand, the 1T$'$ phase of MoTe$_2$, where both superconducting and higher-order topological bands coexist, falls in the middle.
In the presence of a superconducting coherence in the momentum space, the interband interaction between the bulk $s$-wave superconducting and the topological bands results in a proximity-induced $p$-wave superconductivity at the surface, the schematics of which are illustrated in Fig.~\ref{fig1:crys_struct}(b,c).
Therefore, MoTe$_2$ can provide a distinct route to realize intrinsic topological superconductivity overcoming the disadvantages of the other two implementations, similar to suggested topological superconductivity in Fe-based superconductors~\cite{zhang2018observation,qin2023two}.

Despite its potential, studies of the interplay between higher-order topology and superconductivity in MoTe$_2$ are limited, primarily due to the extremely high-pressure environment required to stabilize the 1T$'$ phase, posing significant challenges for ARPES and STM measurements.
Therefore, alternative experimental methods and controlling relevant symmetries are critical to advancing our understanding of the SC properties in the 1T$'$ phase of MoTe$_2$.

Here, we investigate the proximity-induced surface $p$-wave superconductivity and higher-order topological nature of MoTe$_2$ in its high-pressure 1T$'$ phase. Using surface-sensitive soft-point-contact Andreev reflection (PCAR) spectroscopy under quasi-hydrostatic pressure, we confirm the emergence of surface $p$-wave superconductivity by the analysis of the Blonder-Tinkham-Klapwijk (BTK) model and the observation of a zero-bias conductance peak.
As summarized in Fig.~\ref{fig1:crys_struct}, the $p$-wave superconductivity arises through the proximity effect between bulk $s$-wave superconductivity and helical Dirac surface states, which originate from second-order topological insulator (SOTI) bands protected by time-reversal and inversion symmetry.
This explains the similar temperature and magnetic-field dependence for both $s$-wave and $p$-wave gaps, as the surface $p$-wave pairing reflects its proximity-induced nature.
In contrast to intrinsic $p$-wave superconductivity, the observed $p$-wave pairing is easily suppressed by external magnetic fields or when inversion symmetry is broken under reduced pressure, alongside the reduction of the bulk $s$-wave superconductivity.
Furthermore, our analysis, supported by a Bogoliubov-de-Gennes Hamiltonian, suggests that the topological hinge states in the SOTI phase may evolve into zero-energy Majorana corner states in the third-order topological superconducting (TOTSC) phase.
These findings establish MoTe$_2$ as a promising candidate for realizing topological superconductivity and Majorana states, providing a platform for exploring exotic quantum states in TMDC materials.

\begin{figure}[t]
\includegraphics[width=0.7\textwidth]{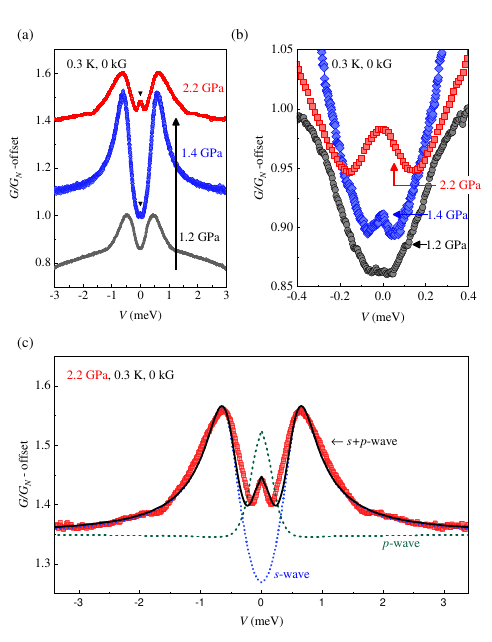}
\caption{\label{fig2_evolofzbcp} 
\textbf{(a)} Normalized differential conductance in the T$_d$ (1.2~GPa) and 1T$'$ (1.4 and 2.2~GPa) phases at 0.3~K. The zero-bias conductance peak (ZBCP), indicated by the black triangle, emerges in the 1T$'$ phase.
\textbf{(b)} Magnified view near the ZBCP of (a).
\textbf{(c)} 
Normalized differential conductance measured at 2.2~GPa.
The blue, green, and black lines represent the $s$-, $p$-, and $s+p$-wave superconductivity, respectively, which are derived from the Blonder-Tinkham-Klapwijk (BTK) fittings of the two-gap $s+p$-wave model.
}
\end{figure}

~

\begin{figure}[t]
\includegraphics[width=\textwidth]{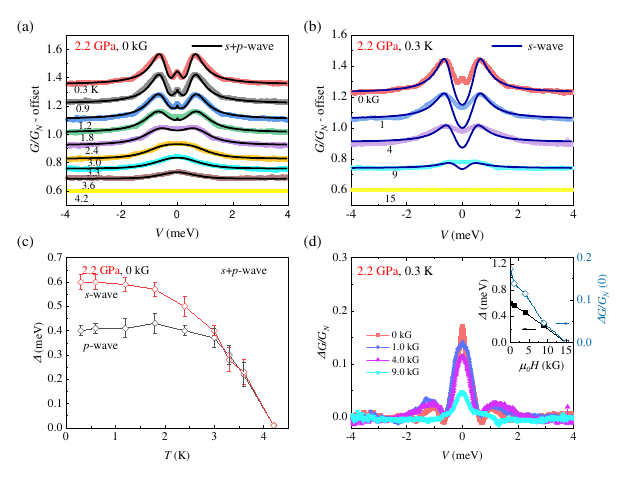}
\caption{\label{fig3_1T'pase} 
\textbf{(a)} Temperature evolution of the normalized differential conductance at 2.2~GPa.
Solid lines are the results for BTK fittings of the two-gap $s+p$-wave model, and colored lines are measured data sets.
\textbf{(b)} Normalized differential conductance under various magnetic fields at 2.2~GPa and 0.3~K. 
ZBCP is suppressed as the magnetic field increases.
Solid lines represent the BTK fitting results based on the $s$-wave model, serving as a guideline to illustrate the suppression of the $s$-wave gap under magnetic fields.
In (a,b), each curve is shifted by a constant offset for clarity.
\textbf{(c)} 
Temperature-dependent superconducting gaps for $s$- and $p$-wave.
\textbf{(d)} The differential conductance after subtracting the $s$-wave's contribution  from (b) for various magnetic fields.
The magnetic field suppresses the ZBCP by increasing the magnetic field.
The inset summarizes the magnetic-field dependence of the magnitude of the ZBCP (open diamond) and the $s$-wave gap (black square) obtained from the data in (b).
}
\end{figure}
\clearpage

\section{Results} \label{sec:results}



We initially investigate the superconductivity of the orthorhombic T$_d$ phase under relatively low pressure to compare with emerging superconductivity in the 1T$'$ phase.
Figure~S2(a,b) shows the temperature- and magnetic-field-dependent PCAR spectra at 0.7 GPa, respectively.
Even though a single-gap $s$-wave superconductivity seems to be enough to explain the temperature-dependent PCAR spectra in Fig.~S2(a), the magnetic-field-dependent PCAR spectra and fitting analysis in Fig.~S2(b-d) clearly reveal the emergence of  $s$-wave superconductivity with two-gap feature below $T_c$. 
Figure~S3 further shows the PCAR data in higher pressure conditions (i.e., 1.2 GPa), indicating the robustness of the two-gap feature.
These results are consistent with the previously reported topologically nontrivial two-gap superconductivity under ambient pressure~\cite{guguchia2017signatures,naidyuk2018surface}.
See the more detailed analysis in Sec.~S1.2 in the Supplementary Information (SI).~\cite{guguchia2017signatures,naidyuk2018surface}.


As pressure increases more, MoTe$_2$ undergoes a structural phase transition from the orthorhombic T$_d$ to the monoclinic 1T$'$ phases and also exhibits a superconductivity below $T_c$ [Fig.~\ref{fig1:crys_struct}(a)].
Figure~\ref{fig2_evolofzbcp}(a) displays the pressure-dependent PCAR spectra for superconducting phases at 0.3 K, ranging from T$_d$ phase (1.2~GPa) to the 1T$'$ (2.2~GPa). 
The data reveal that a small zero-bias conductance peak (ZBCP) is developed within the double conductance peaks.
This feature can be seen clearly in Fig.~\ref{fig2_evolofzbcp}(b), which focuses on the PCAR spectra near the ZBCP. 
Such a ZBCP consistently appears in the temperature-dependent PCAR spectra below $T_c$ at a fixed pressure of 2.2~GPa, as shown in Fig.~\ref{fig3_1T'pase}(a), implying that the ZBCP is an intrinsic feature, as elaborated below.

To investigate the nature of the superconducting gap in the 1T$'$ phase, we conduct the following comprehensive BTK fitting analysis, which indicates that the measured PCAR spectra are most consistently explained by a superconducting state with a combined $s+p$-wave gap.
Above all, the isotropic $s$-wave with a full gap fails to account for the anomalous ZBCP observed between the double conductance peaks, as shown in Fig.~S4(a).
Next, anomalous ZBCPs may arise from Andreev-bound states in exotic superconductors with anisotropic $s$-wave, $p$-wave, or $d$-wave pairing symmetries featuring nodes~\cite{kashiwaya2000tunnelling,deutscher2005andreev}. 
However, an anisotropic $s$-wave scenario is improper due to the ill-fitting, as shown in Fig.~S4(a).
The one-gap $p$-wave or one-gap $d$-wave BTK model could lead to a ZBCP, as depicted in Figs.~S4(b) and S6, with junction angles of $\alpha=\pi/1.35$ and $\alpha=\pi/14$, respectively.
However, such fittings are obsolete due to the condition of the experimental setup being $\alpha=0$.
Furthermore, taken together with the fact that $d$-wave superconductivity typically occurs in a tetragonal crystal structure~\cite{wenger1993d}, these results indicate that $d$-wave superconductivity is unlikely in the monoclinic 1T$'$ phase of MoTe$_2$.
Therefore, only a variant of $p$-wave superconductivity remains as a candidate.

Figure~\ref{fig3_1T'pase}(a) provides compelling evidence that the temperature-dependent PCAR spectra are well-captured by the $s+p$-wave BTK model (black lines), indicating the coexistence of $s$-wave and $p$-wave superconducting gaps. 
This is further supported by Fig.~\ref{fig2_evolofzbcp}(c), which shows the normalized conductance near the ZBCP at 0.3~K, highlighting the distinct contributions of the $s$-wave and $p$-wave components, with the superconducting gaps measured as $\Delta_s = 0.6$~meV and $\Delta_p = 0.4$~meV.
These values correspond to $2\Delta_s/k_B T_c \approx 3.3$ and $2\Delta_p/k_B T_c \approx 2.2$, respectively.
The comparable magnitudes of these gaps strongly suggest that the $p$-wave pairing arises from a robust physical mechanism rather than being an artifact of the experimental setup. 
Detailed fitting parameters for Figs.~\ref{fig3_1T'pase}(a) and \ref{fig2_evolofzbcp}(c) are listed in Table~S4.

To address the intrinsic origin of the ZBCP, we rule out scenarios attributing it to external artifacts by analyzing its response to a magnetic field.
One potential source of the ZBCP is the thermal effect from point-contact heating.
In this case, the ZBCP is generally broad and exhibits sensitivity to weak magnetic fields due to the critical current and high normal-state resistivity
~\cite{wang2019tunability,gorski2018interplay,eiling1981pressurepp}.
To exclude this possibility, we obtain the magnetic field-dependent PCAR spectra at 2.2~GPa and 0.3~K [Fig.~\ref{fig3_1T'pase}(b)].
The experimental data in Fig.~\ref{fig3_1T'pase}(b) does not demonstrate any aforementioned characteristics indicative of the thermal regime, ruling out the contribution of heating effects to the ZBCP.
Another scenario is the presence of in-gap bound states induced by magnetic impurities, which can result in symmetric or asymmetric double conductance peaks in PCAR and a ZBCP under specific conditions ~\cite{manna2020signature,chen2017quantum,wang2016observation,aggarwal2016unconventional}.
If this were the case, the ZBCP would split into two peaks in the presence of a magnetic field, owing to the Zeeman effect. 
%
In Fig.~\ref{fig3_1T'pase}(d), the ZBCP observed at 2.2 GPa remains as a single peak under the magnetic field, thus excluding the magnetic impurity scenario.
These discoveries imply that the ZBCP arises from intrinsic SC properties.

We now investigate the properties of the emerging $p$-wave superconductivity in the 1T$'$ phase.
First, the $p$-wave superconductivity is closely tied to bulk $s$-wave superconductivity via the proximity effect, as illustrated in Fig.1. 
Experimental evidence demonstrates that $s$- and $p$-wave superconducting gaps exhibit strikingly similar temperature-dependent behaviors, as shown in Fig.~\ref{fig3_1T'pase}(c).
Both gaps decrease with increasing temperature and vanish at the same $T_c$, indicating their intrinsic connection.
Moreover, the $s$- and $p$-wave superconductivity also exhibit similar magnetic field responses. 
For this, we distinguish the contribution of $s$- and $p$-wave superconductivity.
Figure~\ref{fig3_1T'pase}(b) depicts the magnetic field-dependent PCAR spectra, which are fitted using a single $s$-wave BTK model (solid lines), implying the suppression of s-wave superconductivity under a magnetic field.
On the other hand, Fig.~\ref{fig3_1T'pase}(d) represents residual $p$-wave contributions after removing the $s$-wave contribution.
Finally, the inset in Fig.~\ref{fig3_1T'pase}(d) summarizes a magnetic field-dependent $s$-wave superconducting gap (solid square) and the magnitude of the ZBCP (open diamond), the latter of which is a hallmark of $p$-wave superconductivity.
They are gradually suppressed by the magnetic field and completely suppressed near 15~kG. 
Unlike intrinsic spin-triplet $p$-wave superconductivity, which is robust against magnetic fields, the observed $p$-wave pairing vanishes more easily, further confirming its induced nature.
These observations indicate that the $p$-wave feature is strongly correlated with the $s$-wave superconductivity, irrespective of tuning factors such as temperature and magnetic field.

It is noteworthy that the role of symmetry is critical in stabilizing the $p$-wave superconductivity. The ZBCP observed in the inversion-symmetric 1T$'$ phase disappears in the inversion symmetry-broken T$_d$ phase, as shown in Figs.~\ref{fig2_evolofzbcp}(a), S2, and S3. 
Also, the magnetic field easily breaks the $p$-wave superconductivity, as shown in Fig.~\ref{fig3_1T'pase}(b,d).  
This indicates that inversion and time-reversal symmetries play a fundamental role in protecting the emerging $p$-wave superconductivity, which is in agreement with the theoretical prediction below.

We now theoretically investigate the emergence of the $p$-wave superconductivity, which will be attributed to the bulk-to-surface proximity effect between $s$-wave bulk superconductivity and helical Dirac-type surface states generated from a second-order topological band.
For this end, we examine a Bogoliubov-de-Gennes (BdG) Hamiltonian and the higher-order nature of the 1T$'$ phase. 
Since the normal state of the 1T$'$ phase is monoclinic and the superconducting phase transition does not alter the system's structure, both the SOTI and BdG Hamiltonians retain inversion and time-reversal symmetries, ensuring PT symmetry [see detailed Hamiltonians in Methods].

\begin{figure}[t]
\centering
\includegraphics[width=0.95\textwidth]{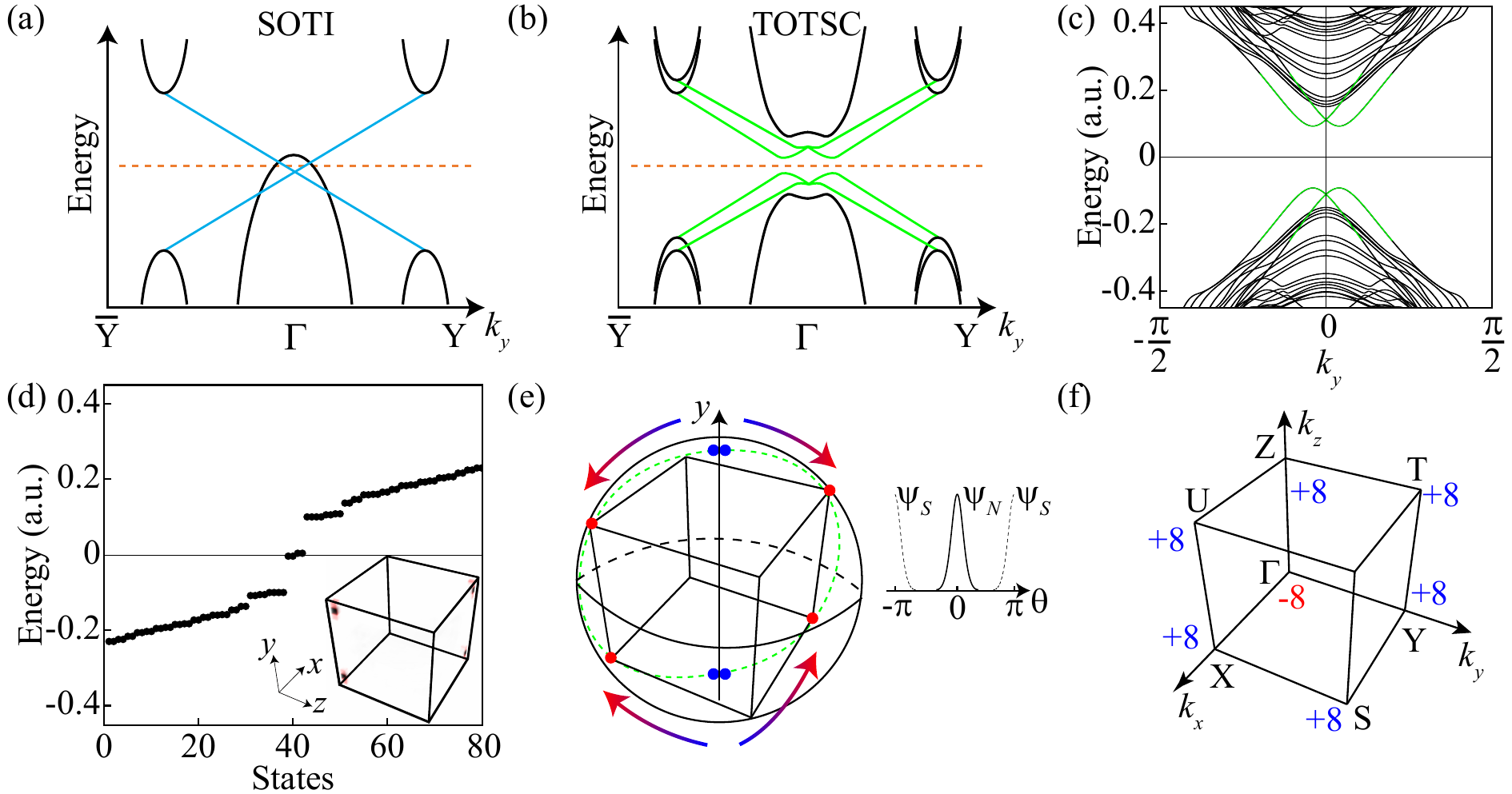}
\caption{\label{fig4_theory}
Schematic band diagrams for \textbf{(a)} the normal and \textbf{(b)} the $s+p$-wave third-order topological superconducting phases.
For the normal phase, the multiband comprises the second-order topological insulating bands and the hole band at the $\Gamma$ point.
The topological hinge states (blue lines) from the SOTI, present in (a), are transformed into gapped hinge states (green lines) of the TOTSC in (b).
In the superconducting gap, Majorana corner states can emerge.
In (a,b), the dashed orange line indicates the position of the zero energy.
\textbf{(c)}
Band structures for the finite tight-binding Hamiltonian composing $15\times15$ unit cells in $xz$ plane and periodic in $y$ (black lines) for the SC phase.
The analytically obtained dispersion relation of the hinge (green line) overlaps with the numerically calculated result.
\textbf{(d)}
Energy eigenvalues for the finite tight-binding Hamiltonian with $15 \times 15 \times 15$  lattices, where there exist four nearly zero-energy states.
The inset shows the wavefunction distributions of the nearly zero-energy corner states, where darker color indicates higher density.
\textbf{(e)} (Left) Schematics of the distribution of the Majorana zero-energy states.
The zero-energy Majorana states at the poles (blue) in spherical geometry move to the corners (red) along the dotted green line when the geometry becomes cubic.
(Right) Schematics of the localized wavefunctions of Majorana corner states along a hinge passing through the north and south poles.
At $\theta=0$ and $\theta=\pi$, the north and south Majorana corner states indicated by the solid and dashed lines, $\Psi_N$ and $\Psi_S$, respectively, are well-localized.
\textbf{(f)}
Parity eigenvalues for the occupied states at the TRIM points.
The parameters are $t_x=t_z=1, t_y=2, \lambda_x=\lambda_z=1,\lambda_y=0.5,m_0=-3,m_1=m_2=0.3,\Delta_{s1}=\Delta_{p1}=0, \Delta_{s2}=0.1, \Delta_{p2}=0.05, \Delta_{p3}=0.04$ and $\mu=0.05$.
}
\end{figure}

The simplified band structure of the normal state is composed of a hole band at the $\Gamma$ point and a SOTI band, as illustrated in Fig.~\ref{fig4_theory}(a), which is derived from the density functional theory (DFT) results~\cite{wang2019higher,huang2024hybrid}.
The surface states arising from the SOTI band consist of double copies of a topological helical surface Dirac cone with spin-momentum-locking~\cite{wang2019higher},
described by the effective surface Hamiltonian $H_{\text{Surf}} \approx \tau_z\left(k_x\sigma_y-\lambda_y k_y\sigma_x\right)$.
Additional Dirac mass terms in the SOTI band render the surface Dirac cones gapless only at the hinges, giving rise to topological hinge states (indicated by blue lines of Fig.~\ref{fig4_theory}(a)), a hallmark feature of the SOTI~\cite{schindler2018higher}.
The SOTI band and its hinge states are reproduced numerically and analytically, as shown in Fig.~S6. 
Further details, including the Hamiltonian, symmetry properties, and numerical and analytical analysis, are provided in Sec.~S2.1 of SI.
%


Considering the Fermi surface generated from the parabolic hole band near the $\Gamma$-point [Fig.~\ref{fig4_theory}(a)], we consider a conventional $s$-wave spin-singlet pairing:
\begin{eqnarray}\label{eq:s-wave}
    H_{\text{Bulk Pairing}} = \Delta_{s\text{-wave}}\xi_x\tau_x\mu_z\sigma_y,
\end{eqnarray}
where the Pauli matrices $\xi_i$ indicates the Nambu space, $\tau_j$ and $\mu_k$ indicate the orbitals, $\sigma_l$ represents the spin of the system, and $\Delta_{s\text{-wave}}$ indicates the $s$-wave pairing gap.
With this pairing, the $s$-wave superconducting Hamiltonian for the bulk parabolic hole band is given by Eq.~(\ref{eq:SC}) in Methods.


To investigate the proximity effect between the SOTI Hamiltonian in Eq.~(\ref{eq:SOTI}) and the $s$-wave superconducting Hamiltonian in Eq.~(\ref{eq:SC}), we introduce the following interaction Hamiltonian:
\begin{eqnarray}\label{eq:int}
    H_{\text{I}}=\sum_{j}\lambda\left(a^\dagger_{j,s}c_{j,s}+c^\dagger_{j,s}a_{j,s}\right),
\end{eqnarray}
where $a_{j,s}^\dagger$ indicates the creation operators for the $j$th orbital with spin $s$ for the parabolic band and $c_{j,s}^\dagger$ does for the SOTI band, and $\lambda$ indicates the magnitude of the coupling~\cite{stanescu2010proximity}.
This interaction facilitates the transfer of bulk $s$-wave superconductivity, described by Eq.~(\ref{eq:s-wave}), to the topological surface states originating from the SOTI band, resulting in the emergence of $p$-wave superconductivity. 
This phenomenon, illustrated in Fig.\ref{fig1:crys_struct}, is referred to as the bulk-to-surface proximity effect~\cite{zhang2018observation,qin2023two}.

To explicitly determine the induced SC pairing via the bulk-to-surface proximity effect, we first derive the effective superconducting Hamiltonian for the SOTI up to the second-order using the Schrieffer-Wolff transformation~\cite{schrieffer1966relation}.
This results in two $s$-wave and three $p$-wave pairings in the bulk system, with the corresponding fermion bilinear forms summarized in Table~\ref{table:pairing}.
Further projecting the bulk Hamiltonian onto the cylindrical surface using surface-localized states $\ket{\psi_j}$, we derive the explicit surface Hamiltonian, $\left[H^\text{BdG}_\text{Surf}(\textbf{k})\right]_{ij}=\bra{\psi_i}
H^{\text{BdG}'}_{\text{SOTI}} (\textbf{k})\ket{\psi_j}$, which is given in Eq.~(\ref{eq:BdGsurface}) in Methods.
Then, the induced surface superconductivity is given as:
\begin{eqnarray}
    \label{eq:induced_Hamiltonian}
    H_\text{Induced} (\textbf{k}) &=&\Delta_{s1}(\sin\phi+\cos\phi)\xi_y\tau_y+\Delta_{s2}\xi_y\tau_y\left(\sin\phi\sigma_x-\cos\phi\sigma_z\right)\\
    &&-\Delta_{p3}\sin k_y\xi_x\tau_x\left(\sigma_0-\sin^2\phi\sigma_x+\sin\phi\cos\phi\sigma_z\right),\nonumber
\end{eqnarray}
where $\xi_i,\tau_j$ and $\sigma_k$ indicate the Nambu space, orbital, and spin degrees of freedom, respectively.
$\phi$ is the polar angle along the cylindrical surface.
This resulting surface superconductivity implies that among the five bulk pairings in Table~\ref{table:pairing}, both $s$-wave ($\Delta_{s1},\Delta_{s2}$) and one $p$-wave ($\Delta_{p3}$) pairings survive on the surface, while the other two $p$-wave pairing terms vanish.
For simplicity, we set $\Delta_{s1}=\Delta_{p1}=0$, as this does not affect the topological index of the system described in Eq.~(\ref{eq:Index}).
Detailed information including nonzero $\Delta_{s1}$ and $\Delta_{p1}$ case are provided in Sec.~S2.2 of SI.

Therefore, we theoretically demonstrate that both $s$- and $p$-wave surface superconductivity are induced by the bulk $s$-wave superconductivity, as illustrated in Fig.~\ref{fig1:crys_struct}. 
Notably, the $p$-wave pairings induced in the SOTI arise from the interplay between the trivial bulk $s$-wave superconductivity and the topological surface states, which are protected by inversion and time-reversal symmetries.
Thus, if the bulk $s$-wave gap, $\Delta_{s\text{-wave}}$, vanishes, no proximity-induced pairings can exist as all the induced bulk pairings are proportional to $\Delta_{s\text{-wave}}$ (see detailed form in Sec.~S2.2.2 of SI).
This theoretical analysis aligns with the experimental results in Figs.~\ref{fig2_evolofzbcp} and \ref{fig3_1T'pase}, where both the $p$-wave and $s$-wave pairings vanish at the same temperature and are fragile under inversion symmetry breaking or an applied magnetic field.

\begin{table}[t]
\begin{tabular}{ p{1.5cm} p{10cm} p{4cm}}
\hline
$s$-wave & Fermion Bilinear & Matrix Form \\
\hline
$\Delta_{s1}$ & $i\sum_{j=1}^2 \left[-c_{j,\uparrow}^\dagger c_{j+2,\downarrow}^\dagger+c_{j,\downarrow}^\dagger c_{j+2,\uparrow}^\dagger\right]$ & $\xi_x\left(\tau_x+\tau_y\mu_y\right)\sigma_y$ \\
&
$+i\sum_{k=1}^2 (-1)^{k-1}\left[c_{k,\uparrow}^\dagger c_{5-k,\downarrow}^\dagger-c_{k,\downarrow}^\dagger c_{5-k,\uparrow}^\dagger\right]$+h.c.&\\
$\Delta_{s2}$ & $i\sum_{j=1}^2 (-1)^j\left(c_{j,\uparrow}^\dagger c_{j+2,\downarrow}^\dagger-c_{j,\downarrow}^\dagger c_{j+2,\uparrow}^\dagger\right)$+h.c. & $\xi_x\tau_x\mu_x\sigma_y$\\
\hline\hline
$p$-wave & Fermion Bilinear & Matrix Form \\
\hline
$\Delta_{p1}$ & $i\left(\sin k_x-\sin k_z\right) \sum_{j}(-1)^{j-1} c_{j,\uparrow}^\dagger c_{j+2,\downarrow}^\dagger-c_{j,\downarrow}^\dagger c_{j+2,\uparrow}^\dagger$+h.c. & $\left(\sin k_x-\sin k_z\right)\xi_y\mu_y\sigma_y$\\
$\Delta_{p2}$ & $\sin k_y \sum_{j}c_{j,\uparrow}^\dagger c_{j,\downarrow}^\dagger+c_{j,\downarrow}^\dagger c_{j,\uparrow}^\dagger$+h.c. & $\sin k_y\xi_x\sigma_x$ \\
$\Delta_{p3}$ & $\sin k_y \sum_{j}(-1)^{\left[\frac{j-1}{2}\right]_\text{floor}}c_{j,\uparrow}^\dagger c_{j,\downarrow}^\dagger+c_{j,\downarrow}^\dagger c_{j,\uparrow}^\dagger$ & $\sin k_y\xi_x\left(\tau_z+\mu_x\right)\sigma_x$\\
 & $+\sin k_y \sum_{k=1}^2c_{2k-1,\uparrow}^\dagger c_{2k,\downarrow}^\dagger+c_{2k-1,\downarrow}^\dagger c_{2k,\uparrow}^\dagger$+h.c.& \\
\hline
\end{tabular}
\caption{\label{table:pairing}
\textbf{Induced $s$- and $p$-wave pairings in the bulk system.}
The pairings are expressed by their Fermion bilinear and respective matrix forms.
In the column of Fermion bilinear, the index $j$ indicates the orbital, with the possible index being $j=1,2,3,4$.
The $16\times16$ matrix form for the pairing functions is expressed by the tensor product of Pauli matrices $\Delta(\textbf{k}) \xi_i \tau_j \mu_k \sigma_l$ except for $\Delta_i$.
Here, $\Delta_i$ is the coefficient for the superconducting gap function, the Pauli matrices $\xi_i$ indicate the Nambu space, $\tau_j$ and $\mu_k$ indicate orbitals, and $ \sigma_l$ indicate the spin degrees of freedom.
}
\end{table}

\clearpage

\clearpage


Finally, we explore the potential emergence of Majorana corner states through numerical and analytical methods. 
We begin by examining the evolution of the topological hinge state in the SOTI.
The topological hinge states (blue lines) in the SOTI, as shown in Fig.~\ref{fig4_theory}(a), transform into gapped hinge states (green lines) in the TOTSC, as depicted in Fig.~\ref{fig4_theory}(b).
Figure~\ref{fig4_theory}(c) presents the calculated band structure for the square geometry in the $xz$-plane and periodic boundary conditions along the $y$-axis ($C_2$ symmetry axis).
The band structure is obtained through both numerical and analytical approaches.
Numerically, it is derived using a finite tight-binding (TB) BdG Hamiltonian, while
analytically, the Jackiw-Rebbi method~\cite{jackiw1976solitons} is employed.
Both approaches consistently demonstrate that the hinges of the SOTI bands become gaped as the system transitions into a TOTSC, as illustrated in Fig.~\ref{fig4_theory}(a,b)~\cite{Supple}.
As shown in Fig.~\ref{fig4_theory}(c), the analytical dispersion (green lines) closely matches the numerical dispersion from the TB calculation (black lines), confirming the gapped hinge states. 
This gapped hinge system effectively provides a one-dimensional topological superconductor along the hinge, supporting the Majorana corner state at its termination.

To see the Majorana corner states of the TOTSC, we calculate the energy spectrum of a finite system with a cubic geometry using the TB Hamiltonian [Fig.~\ref{fig4_theory}(d)], where four nearly zero-energy in-gap states are observed.
Note that the nearly zero-energy in-gap state is converging to zero energy with increasing system size, and it is due to the finite size effect of the numerical calculation while its energy is at exact zero analytically, which can be seen in Secs.~S2.2.5 and S2.2.7 of SI.
The spatial distribution of the wavefunction associated with each zero-energy state confirms its localization at the corners of the cubic geometry, as depicted in the inset of Fig.~\ref{fig4_theory}(d)
The emergence of these zero-energy Majorana states can be further investigated using the spherical geometry.
The zero-energy Majorana states of the system are localized near $\theta=0$ and $\pi$ of the polar angle $\theta$ in respect to the $y$-axis [Fig.~\ref{fig4_theory}(e)].
The four localized wavefunctions are at the poles of the spherical geometry and descend, each equally splitting into two corners for the cubic geometry.
The schematics can be seen in Fig.~\ref{fig4_theory}(e).

The resulting zero-energy Majorana corner states arising in cylindrical and spherical geometry form Majorana Krammers' pairs~\cite{haim2019time,zhang2013time}, with their wavefunctions denoted as $\psi^{\pm}_{j}$ for $j = T, B, N, S$.
These wavefunctions are localized at the top and bottom (north and south poles) of the system $\psi^\pm_{T,B}$ $\left(\psi^\pm_{N,S}\right)$.
They satisfy the particle-hole symmetry
    $C\psi^\pm_{T,B,N,S}= \pm \psi_{T,B,N,S}^\pm$
and time-reversal symmetry
    $\Theta\psi^\pm_{T,B,N,S}= \mp i \psi_{T,B,N,S}^\mp$
for the charge-conjugation and time-reversal operators $C=\xi_x K,\Theta=i\sigma_y K$, where $K$ is a conjugation operator.
It can be seen that $\left(\psi_{T,B,N,S}^{\pm}\right)^C=\pm\psi^{\pm}_{T,B,N,S}$, indicating that the wavefunctions are indeed Majorana Krammers' pairs.
Thus, any combination of the Majorana pairs cannot create either a wavefunction of pure particle or hole.
See the detailed analysis regarding the Majorana corner states in Method and SI.

The BdG Hamiltonian of this system, with redundant particle-hole symmetry, falls within the DIII class of the Altland-Zirnbauer classification, confirming that topological Majorana boundary states can emerge~\cite{schnyder2008classification, khalaf2018higher, huang2024hybrid}.
The topological nature of the corner states can be found using a $\mathbb{Z}_8$ index given as~\cite{ahn2020higher,khalaf2018symmetry,huang2024hybrid}
\begin{eqnarray}\label{eq:Index}
    \kappa=\frac{1}{4}\sum_{K\in \text{TRIM}}\left[N^+(K)-N^-(K)\right]~\text{mod}~8=4,
\end{eqnarray}
where $N^\pm(K)$ indicates the number of occupied eigenstates with parity eigenvalues $\pm1$ at the TRIM point $K$.
The parity eigenvalues for each occupied state for each TRIM point can be seen in Fig.~\ref{fig4_theory}(f).
The topological index $\kappa$ indicates that the system is trivial when $\kappa=0$, a first-order topological system when $\kappa=1,3,5,7$, a second-order topological system when $\kappa=2,6$, and a third-order topological system when $\kappa=4$, which explains the existence of the zero-energy corner states.
Note that the total Hamiltonian in Eq.~(\ref{eq:Htot}), composed of SC and SOTI systems, and the proximity-effect-induced superconducting SOTI system in Eq.~(\ref{eq:BdGHamiltonian}) share the same topological invariant, $\kappa=4$, regardless of Schrieffer-Wolff transformation, as shown in Fig.~S8 (see details in Sec.~S2.2 of SI).
Therefore, our theoretical analysis and experimental results indicate that symmetry-protected Majorana corner states can emerge in the proximity-induced surface SC state of the 1T$'$ phase.

\clearpage

\section{Conclusion}
We investigated superconductivity in both the T$_d$ and 1T$'$ phases of MoTe$_2$ through surface-sensitive soft-point contact Andreev reflection (PCAR) spectroscopy measurements under extreme conditions.
In the orthorhombic T$_d$ phase, an $s$-wave SC state with two-gap features was observed, which is consistent with the previously reported topologically nontrivial superconductivity~\cite{guguchia2017signatures}. 
As pressure increases, MoTe$_2$ undergoes a phase transition from the orthorhombic T$_d$ to the monoclinic 1T$'$ phase. 
In the 1T$'$ phase, a zero-bias conductance peak appears between the double conductance peaks in the PCAR spectra, where the extended BTK model with the coexistence of $s$- and $p$-wave pairings consistently explains the experimental results.
The emergence of the surface $p$-wave pairing in the 1T$'$ phase is attributed to the bulk-to-surface proximity effect between the $s$-wave superconductivity of the bulk and the topological helical Dirac surface state, which is consistent with the higher-order topological nature of the normal phase.
Furthermore, our theoretical analysis suggests that the topological hinge states in the second-order topological insulating (SOTI) phase may evolve into Majorana corner states in the third-order topological superconducting (TOTSC) phase, with the $p$-wave superconducting gap playing a key role in their emergence.

\newpage
\section{Methods}
\subsection{Experimental Methods}
Under quasi-hydrostatic pressurized conditions, MoTe$_2$ undergoes a phase transition from the T$_d$ to 1T$'$ phases.
To create quasi-hydrostatic pressure environments, the samples are enclosed within a Teflon tube and pressurized using a clamp-type hybrid cell filled with Daphne 7373 as the pressure-transmitting medium.
This setup allows for the application of quasi-hydrostatic pressures of up to 2.7~GPa.
The pressure inside the cell at low temperatures is determined by monitoring the sharp resistivity drop at the SC transition temperature of Pb and Sn~\cite{Walker1999pressure,eiling1981pressurepp}.

Soft point-contact spectroscopy is a highly sensitive technique utilized for probing the gap structure of superconductors and the surface properties of materials~\cite{naidyuk2018surface}.
In our study, we applied this technique to investigate the gap structure of superconductivity in MoTe$_2$ across the phase transition from T$_d$ to 1T$'$. The point contact was established using silver grains deposited on the sample surface within a quasi-hydrostatic pressure environment.
After screening numerous crystals, the optimal conditions for soft point-contact spectroscopy during pressure experiments were identified.
In several experiments, the contacts remained in the ballistic regime under high-pressure conditions without transitioning to the thermal regime.
Temperature regulation down to 0.25~K was achieved using the $^3$He Heliox System (Oxford Instruments Nanotechnology).

\subsection{Theoretical Methods}
For the normal state of MoTe$_2$ in the 1T$'$ phase, we consider a simplified TB model describing the SOTI bands, incorporating additional higher-order spin-orbit coupling terms to align with DFT calculations~\cite{wang2019higher,ezawa2019second,huang2024hybrid}.
The TB Hamiltonian $H _{\text{SOTI}}(\textbf{k})$ for the normal state of SOTI bands is given by
\begin{eqnarray}\label{eq:SOTI}
    H_{\text{SOTI}}(\textbf{k}) &=&
    \left(m_0+\Sigma_{i=x,y,z}t_i \cos k_i\right) \tau_z+ \lambda_x \sin k_x \tau_y\mu_y+\lambda_z \sin k_z \tau_x\\
    &&+\lambda_y \sin k_y\tau_y\mu_z\sigma_z+m_1 \tau_z\mu_z+m_2\mu_x    \nonumber\,
\end{eqnarray}
where the Pauli matrices $\tau_i$ and $\mu_j$ represent the orbital degrees of freedom and $\sigma_k$ the spin degrees of freedom, respectively.
$t_i$ are simple hopping terms between the nearest neighbors while
$\lambda_j$ are either spin- or orbital-coupled hopping terms.
$m_1$ and $m_2$ are symmetry invariant mass terms.
For the emergence of the SOTI hinge states, we choose $m_1=m_2=m$ without loss of generality.

For an $s$-wave superconducting band near $\Gamma$ in the bulk of the system, we consider the following low-energy effective Hamiltonian:
\begin{eqnarray}\label{eq:SC}
    H_\text{SC} (\textbf{k})=\left(-k_y^2+\varepsilon_0\right)\xi_z+\Delta_{s\text{-wave}} \xi_x\tau_z\mu_z\sigma_y,
\end{eqnarray}
where the Pauli matrices $\xi_i$ represents the Nambu space.
$\varepsilon_0$ is the constant energy shift.

To consider the proximity effect between the SOTI and bulk $s$-wave superconductor, we consider the interaction Hamiltonian in Eq.~(\ref{eq:int}).
Then, the total system is composed of the SOTI, SC, and interaction Hamiltonians, which is given as
\begin{equation}
\label{eq:Htot}
H_\text{Tot} (\textbf{k}) =
\begin{pmatrix}
H_\text{SC}(\textbf{k}) & H_{\text{I}} \\
H_{\text{I}}^\dagger & H_\text{SOTI}^\text{BdG} (\textbf{k}) 
\end{pmatrix}.
\end{equation}
Here, $H_\text{SOTI}^\text{BdG} (\textbf{k})$ is the BdG expansion of the SOTI Hamiltonian in Eq.~(\ref{eq:SOTI}).

Using the Schrieffer-Wolff transformation, we obtain the following effective BdG Hamiltonian for the SOTI system (see details in Sec.~S2.2).
\begin{equation}
\label{eq:BdGHamiltonian}
H^{\text{BdG}} (\textbf{k}) =
\begin{pmatrix}
H_\text{SOTI}(\textbf{k}) - \mu & \Delta(\textbf{k}) \\
\Delta^\dagger(\textbf{k}) & -H_\text{SOTI}^T(-\textbf{k}) + \mu 
\end{pmatrix}.
\end{equation}
Here, $\Delta(\textbf{k})$ is the induced SC pairing function listed in Table~\ref{table:pairing}, and $\mu$ is the chemical potential.

To confirm the emergence of $p$-wave superconductivity on the surface, we derive the surface Hamiltonian by projecting the BdG Hamiltonian over the surface localized wavevectors derived from the Dirac equation~\cite{schindler2020dirac}.
Using the surface localized wavefunction $\ket{\psi_i}$, the surface Hamiltonian is given as
\begin{equation}
\label{eq:surface}
\left[H^\text{BdG}_\text{Surf}(\textbf{k})\right]_{ij}=\bra{\psi_i}
H^{\text{BdG}} (\textbf{k})\ket{\psi_j}.
\end{equation}
On the cylindrical surface, the resulting surface Hamiltonian is given as
\begin{eqnarray}
\label{eq:BdGsurface}
H^\text{BdG}_\text{Surf} (\textbf{k})&=&\mu\xi_z-k_\parallel\xi_z\sigma_y-m(\cos\phi-\sin\phi)\xi_z\left(\cos\phi\sigma_x+\sin\phi\sigma_z\right)\\
&&-\lambda_y \sin k_y\tau_z\left(\cos\phi\sigma_z-\sin\phi\sigma_x\right)
-\Delta_{s2}\xi_y\tau_y\left(\sin\phi\sigma_x-\cos\phi\sigma_z\right)\nonumber\\
&&+\Delta_{p3}\sin k_y\xi_x\tau_x\left(\sigma_0-\sin^2\phi\sigma_x+\sin\phi\cos\phi\sigma_z\right)\nonumber,
\end{eqnarray}
where the Pauli matrices $\xi_i,\tau_j$ and $\sigma_k$ indicate the Nambu space, orbital and spin degrees of freedom, respectively.
$\phi$ is the polar angle in the $xz$ plane in the cylindrical geometry, and $k_\parallel$ is the momentum along its circular surface.
See the detailed calculation in Sec.~S2:
The surface projection method is explained in Sec.~S2.1.2, and the detailed calculations for the superconducting system are shown in Sec.~S2.2.3.

The Majorana hinge states and their energy-momentum dispersion relation can be obtained by solving the surface BdG Hamiltonian in Eq.~(\ref{eq:BdGsurface}) using the Jackiw-Rebbi method~\cite{jackiw1976solitons} after a transformation of $k_\parallel\rightarrow-i\partial_\parallel$.
Solving the Jackiw-Rebbi differential equation along the surface with $m(\cos\phi-\sin\phi)$ being the effective Dirac mass term, the dispersion near $\phi = \pi/4$ is given by
\begin{eqnarray}\label{eq:BdGhinge}
    H^\text{BdG}_\text{Hinge} (k_y)&=&-\mu\xi_z+\lambda_y \sin k_y\sigma_z+\Delta_{s2}\xi_y\sigma_y+\frac{\sqrt{2}+1}{\sqrt{2}}\Delta_{p3}\sin k_y\xi_x\sigma_x,
\end{eqnarray}
which shows a gapped energy-momentum dispersion relation.
Solving the Jackiw-Rebbi equation using the Dirac equation in Eq.~(\ref{eq:BdGhinge}), redefining the $s$- and $p$-wave sections as $\Delta_s \equiv \Delta_{s2}$ and $\Delta_p \equiv \frac{\sqrt{2}+1}{\sqrt{2}}\Delta_{p3}$ for simplicity, the wavefunction of the zero-energy top and bottom Majorana corner states $\psi_T^\pm(y)$ and $\psi_B^\pm(y)$ are given as: 
\begin{eqnarray}
\psi_T^\pm(y)=\exp\left[\frac{\Delta_s \lambda_y - \mu\Delta_p}{\lambda_y^2+\Delta_p^2} \left(-y+\frac{L}{2}\right) \right]
\begin{pmatrix}
e^{i\theta_-} \\
\pm e^{-i\theta_+} \\
\pm e^{-i\theta_-} \\
e^{i\theta_+}
\end{pmatrix},
\label{eq:top_Majorana_corner}
\\
\psi_B^\pm(y)=\exp\left[\frac{\Delta_s \lambda_y - \mu\Delta_p}{\lambda_y^2+\Delta_p^2} \left(y+\frac{L}{2}\right) \right]
\begin{pmatrix}
e^{i\theta_+} \\
\pm e^{-i\theta_-} \\
\pm e^{-i\theta_+} \\
e^{i\theta_-}
\label{eq:bottom_Majorana_corner}
\end{pmatrix},
\end{eqnarray}
where $\theta_\pm=\frac{\lambda_y \mu + \Delta_s\Delta_p}{\lambda_y^2+\Delta_p^2}\pm\frac{\pi}{4}$.
The range of $y$ for the finite hinge geometry is $-\frac{L}{2}<y<\frac{L}{2}$ with the corners located at $\pm\frac{L}{2}$.
The wavefunctions $\psi^\pm_T(y)$ and $\psi^\pm_B(y)$ are Majorana pairs, respectively.
It should also be noted that these Majorana pairs cannot create either a wavefunction of pure particle or hole, which is clear when we consider the form of the wavefunctions.
The wavefunctions each satisfy $C\psi_{T, B}^\pm= \pm \psi_{T, B}^\pm$ for the charge-conjugation operator $C=\xi_x K$, where $K$ is a conjugation operator indicating particle-hole symmetry.
Therefore, the wavefunctions satisfy the Majorana condition of $[\psi^\pm_{T, B}]^C=\pm \psi^\pm_{T, B}$.

Using the same methodology for the spherical coordinate, we analytically calculate the Majorana corner states localized at the north and south poles, as shown in Fig.~\ref{fig4_theory}(e). The resulting Jackiw-Rebbi solutions for the Majorana corner states at the north and south poles $\psi_N^\pm(\theta)$ and $\psi_S^\pm(\theta)$ are 
\begin{eqnarray*}
\psi_N^\pm(\theta)=e^{F_+(\theta)}
\begin{pmatrix}
e^{i G_-(\theta)} \\
\pm e^{-iG_+(\theta)} \\
\pm e^{-iG_-(\theta)} \\
e^{iG_+(\theta)}
\end{pmatrix},\\
\psi_S^\pm(\theta)=e^{F_-(\theta)}
\begin{pmatrix}
e^{i G_+(\theta)} \\
\pm e^{-iG_-(\theta)} \\
\pm e^{-iG_+(\theta)} \\
e^{iG_-(\theta)}
\end{pmatrix},
\end{eqnarray*}
where the real functions $F_\pm(\theta)$ and $G_\pm(\theta)$ are given via the polar angle $\theta$ from the $y$-axis.
When $\theta=\pi/2$, these Majorana wavefunctions converge to their cylindrical geometric counterparts in Eqs.~(\ref{eq:top_Majorana_corner}) and (\ref{eq:bottom_Majorana_corner}) [see detailed calculations in Sec.~S2.2.7].
The wavefunctions $\psi_N^\pm(\theta)$ and $\psi_S^\pm(\theta)$ are Majorana pairs, respectively.
The wavefunction $\psi^\pm_{N,S}(\theta)$ localized at a pole in the spherical geometry equally splits into the wavefunctions $\psi^\pm_{T,B}(y)$ along both hinges in the cubic geometry.
As was the case for the cylindrical geometry, the Majorana pairs cannot create either a wavefunction of pure particle or hole.
When considering symmetry, the wavefunctions each satisfy the particle-hole symmetry $C\psi^\pm_{N, S}=\pm\psi^\pm_{N, S}$ for the charge-conjugation operator $C=\xi_x K$.
When examining the wavefunctions, it can be seen that $\left[\psi^\pm_{N, S}(\theta)\right]^C=\pm\psi^\pm_{N, S}(\theta)$, indicating that the wavefunctions are Majorana states.

Further comprehensive information for the theoretical analysis can be found in Sec.~S2.

\section*{Data Availability} 
All data supporting this paper are available within the article and the Supplementary Information. Data used in this paper are available from S. Cheon and T. Park on reasonable request.

\section*{Acknowledgements} \label{sec:acknowledgements}
    We acknowledge a fruitful discussion with Y. Kohama, D. Peterson, Youngkuk Kim, Kee-Su Park, and J. D. Thompson. The work at Sungkyunkwan University was supported by an NRF grant funded by the Ministry of Science and ICT (No.2021R1A2A2010925, RS-2023-00220471, 2022H1D3A01077468). The work at Hanyang University was supported by NRF of Korea through Basic Science Research Programs (Grants No. NRF-2021R1H1A1013517, NRF-2022R1A2C1011646, NRF-2022M3H3A1085772, NRF-2022M3H3A10630744, RS-2024-00416036), and the POSCO Science Fellowship of POSCO TJ Park Foundation. This work was also supported by the Quantum Simulator Development Project for Materials Innovation through the NRF funded by the MSIT, South Korea (Grant No. NRF-2023M3K5A1094813).
    This material is based upon work supported by the US Department of Energy, Office of Science, National Quantum Information Science Research Centers, and Quantum Science Center. S. L. was funded by QSC to perform data analysis and manuscript writing. 
    A portion of this work was performed at the National High Magnetic Field Laboratory, which is supported by National Science Foundation Cooperative Agreement No. DMR-2128556* and the State of Florida.

\section*{Author contributions} \label{sec:Authorcontributions}
S. Lee conducted the high-pressure measurements.
M. Kang and S. Cheon carried out the theoretical analysis.
J. Kim and S. Cho synthesized the bulk samples.
S. Lee, M. Kang, D.Y. Kim, S. Cheon, and T. Park wrote the manuscript with input from all authors.

\bibliographystyle{naturemag} 
\bibliography{Ref}

\end{document}